\documentclass[journal,twoside,web]{ieeecolor}

\usepackage{generic}
\usepackage{cite}
\usepackage{graphicx}
\usepackage{textcomp}

\usepackage{subfigure}
\usepackage{epsfig} 
\usepackage{cite}
\usepackage{lcsys}
\usepackage{amsmath,amssymb,amsfonts}
\usepackage{graphicx}
\usepackage{subcaption}
\usepackage{url}
\usepackage{lineno,hyperref}
\usepackage{textcomp}
\usepackage{wrapfig,lipsum,booktabs}
\usepackage{booktabs}
\usepackage[compatibility=false]{caption}
\usepackage{wrapfig}
\usepackage{mathtools}
\usepackage{graphicx}
\usepackage{mathrsfs}
\usepackage{booktabs}
\usepackage{siunitx}
\usepackage{lipsum}
\usepackage{url}
\usepackage{subfloat}
\usepackage{algorithm}
\usepackage{algorithmicx}
 \usepackage{algpseudocode}
\usepackage{caption}

\newtheorem{remark}{Remark}[section]

\newcommand{\ve}[1]{\boldsymbol{#1}}

\begin{document}

\def\BibTeX{{\rm B\kern-.05em{\sc i\kern-.025em b}\kern-.08em
    T\kern-.1667em\lower.7ex\hbox{E}\kern-.125emX}}
\markboth{\journalname, VOL. XX, NO. XX, XXXX 2017}
{Author \MakeLowercase{\textit{et al.}}: Preparation of Papers for IEEE Control Systems Letters (August 2022)}

\title{Fine-Tuning a Simulation-Driven Estimator}

\author{Braghadeesh Lakshminarayanan, \IEEEmembership{Graduate Student Member, IEEE}, Margarita A. Guerrero, Cristian R. Rojas, \IEEEmembership{Senior Member, IEEE}
\thanks{}
\thanks{This work has been partially supported by the Swedish Research Council under contract number 2023-05170, and by the Wallenberg AI, Autonomous Systems and Software Program (WASP) funded by the Knut and Alice Wallenberg Foundation. The authors are with the Division of Decision and Control Systems, KTH Royal Institute of Technology, 100 44 Stockholm, Sweden (e-mails: blak@kth.se; mags3@kth.se; crro@kth.se).
      }%
      \thanks{{Code available at \href{https://github.com/Braghadeeshln/FTSDE}{https://github.com/Braghadeeshln/FTSDE}}.}
}

\maketitle
\thispagestyle{empty}

\begin{abstract}
Many industries now deploy high-fidelity simulators (digital twins) to represent physical systems, yet their parameters must be calibrated to match the true system. This motivated the construction of simulation-driven parameter estimators, built by generating synthetic observations for sampled parameter values and learning a supervised mapping from observations to parameters. However, when the true parameters lie outside the sampled range, predictions suffer from an out-of-distribution (OOD) error. This paper introduces a fine-tuning approach for the Two-Stage estimator that mitigates OOD effects and improves accuracy. The effectiveness of the proposed method is verified through numerical simulations.
\end{abstract}


\section{Introduction}

Recent advances in high-fidelity simulation have popularized digital twins (DTs)--software replicas of physical systems that let engineers explore operating conditions safely and cheaply before deployment~\cite{Grieves2017,kritzinger2018digital,vanderhorn2021digital}. In practice, once a DT is engineered, most modeling choices are fixed; what remains is calibrating a small set of parameters so the DT replicates a specific physical system.

When a DT is available, one can generate synthetic input–output data~\cite{piga2024synthetic} over a range of parameter settings and train a supervised machine learning model that maps observations to these settings, thereby yielding a \emph{simulation-driven estimator}. At implementation, this simulation-driven estimator takes real measurements and returns parameter estimates~\cite{Diskin,garatti2013new,GHOSH2024111327}. The Two-Stage (TS) estimator~\cite{garatti2013new} is an attractive simulation-driven estimator with a simple structure that involves two stages: (i) first, the observations are compressed into representative features, and then (ii) these compressed features, along with their corresponding parameter values, are used to train a supervised machine learning model. This structure avoids the need for an explicit likelihood function and it requires low computation time~{\cite[App.~B.2]{lakshminarayanan2025asymptotic}} for producing estimates, due to extensive offline training, compared to standard point estimators like the prediction error method (PEM) and the dual Extended Kalman Filter (dual-EKF)~\cite{wan2001dual}. The TS estimator can also be used to learn control design rules, effectively taking the \emph{human out of the loop}~\cite{dettu2024data}.

A limitation of TS is prior-range misspecification: when the parameters of the true system lie outside the range of the synthetic parameter values used during training, the accuracy of the pretrained TS estimator can be low; this is called an out-of-distribution (OOD) case. In this paper, we tackle this issue by fine-tuning only the second stage of TS, implemented as a multi-head neural network for its end-to-end differentiability, using the DT and the new observations.

{Our novel fine-tuning procedure first compares the compressed features from real data to those produced by the DT at the TS initial prediction, to determine whether an OOD case has occurred by applying a statistical test based on a whitened feature discrepancy measure constructed from estimated feature covariances. If an OOD case is detected, we: (i) compute Gauss–Newton (GN) iterations on the initial prediction by TS, (ii) generate a confidence set around the GN update and re-simulate a targeted local training dataset with the DT, and {(iii) fine-tune the final portion of the second stage of the TS estimator, which is inspired from \emph{transfer learning}~\cite{pan2009survey} approaches to system identification~\cite{forgione2023adaptation,niu2022deep}}.} {Our work differs from modern simulation-based inference methods~\cite{zeghal2022neural,deistler2022truncated,ward2022robust}, which aim to infer a full Bayesian posterior. In addition, within these methods, some neural posterior estimation methods use differentiable simulator gradients~\cite{zeghal2022neural}, and others use truncated proposals for sequential refinement~\cite{deistler2022truncated}. In contrast, our approach is a non-Bayesian scheme for a pre-trained point estimator, requiring no differentiable simulator. We perform a lightweight, final-layer update, which is conceptually closer to other transfer learning methods in system identification~\cite{forgione2023adaptation,niu2022deep} and broader OOD adaptation strategies~\cite[Ch. 19-20]{pml1Book}}. In particular, our contributions are the following:
\begin{itemize}
\item We derive a feature-space OOD test based on DT-derived feature statistics and whitening;
\item we propose a label-free fine-tuning approach that combines Gauss–Newton iterations with a confidence set to generate a fresh synthetic dataset for the retraining of TS;
\item we design a minimal update of the second-stage of TS, adjusting only its final layers; and
\item we numerically validate our proposed approach against standard baselines.
\end{itemize}

{The remainder of the paper is organized as follows. In Section~\ref{sec:problem_ts}, we introduce the problem statement. Section~\ref{sec:ts} describes the TS estimator. In Section~\ref{sec: proposed_approach}, our method for fine-tuning the pretrained TS estimator is detailed, and in Section~\ref{sec:simulation_study}, we demonstrate the effectiveness of our approach through several numerical examples. Section~\ref{sec:future} outlines the future work. Finally,  Section~\ref{sec: conclusion} concludes the paper.}

\section{Problem Statement}\label{sec:problem_ts}

{We consider a data-generating mechanism \(\mathcal{M}(\ve{\theta})\), a high-fidelity simulator or DT, parameterized by \(\ve{\theta}\in\Theta\subseteq\mathbb{R}^d\).
For notational simplicity, we assume a single-input single-output (SISO) setting: given an
\(N\)-length \emph{scalar} input sequence \(\ve{u} =(u_1,\dots,u_N)^\top\in\mathbb{R}^N\),
the DT produces a (possibly stochastic) \emph{scalar} output sequence 
\(\ve{y}=(y_1,\ldots,y_N)^\top\in\mathbb{R}^N\).
Concretely, \(\ve{y}=\ve{y}(\ve{\theta};\omega)\), where \(\omega\) denotes the simulator's
noise seed (or any internal randomness). Assume that a length-\(N\) trajectory
\(
\boldsymbol{z}_0=(u_1,y_{0,1},\ldots,u_N,y_{0,N})\in\mathbb{R}^{2N}
\)
has been measured from the DT at some unknown parameter \(\ve{\theta}_0\in\Theta\).
The goal is to estimate \(\ve{\theta}_0\) from \(\boldsymbol{z}_0\) by constructing a \emph{simulation-driven parameter estimator}.
%

{Given a \emph{calibration set} \(\Theta_p\subseteq\Theta\), constructed from prior domain knowledge and possible operating regimes, we draw independent and identically distributed (i.i.d.) parameter settings \(\{\tilde{\ve{\theta}}_i\}_{i=1}^m\) from \(\Theta_p\)}. For each draw $\tilde{\ve{\theta}}_i$, we run the simulator with (possibly) its own noise seed \(\omega_i\) and the chosen input sequence\footnote{Due to space limitations, in this paper the input sequence is considered fixed, even though the treatment can be extended to random inputs.} \(\ve{u}\), obtaining \(\boldsymbol{z}_i=(u_{1},y_{i,1},\ldots,u_{N},y_{i,N})\).
This yields a synthetic, labeled dataset
\(\mathcal{D}=\{(\boldsymbol{z}_i,\tilde{\ve{\theta}}_i)\}_{i=1}^m\). }
An estimator $\hat f_{\mathrm{pre}}\colon \mathbb{R}^{2N}\!\to\mathbb{R}^d$ is then fitted by solving the following minimization problem over a prescribed function class $\mathcal{F}$ (e.g. a neural network):
{
\begin{equation}\label{eq:pre_generic}
\hat f_{\mathrm{pre}} \in
\arg\min_{F\in\mathcal{F}}
\frac{1}{m}\sum_{i=1}^m
L\!\left(F(\boldsymbol{z}_i),\,\tilde{\ve{\theta}}_i\right),
\end{equation}
where $L\colon \mathbb{R}^d \times \mathbb{R}^d \to \mathbb{R}_{\geq 0}$ is a suitable loss function. }
Following Eq.~\eqref{eq:pre_generic}, an estimate of $\ve{\theta}_0$ is obtained as 
\begin{equation}
    \hat{\ve{\theta}}_0 = \hat{f}_{\mathrm{pre}}(\ve{z}_0),
\end{equation}
where $\ve{z}_0$ are the observations (input-output data) collected from the system $\mathcal{M}(\ve{\theta}_0)$. 

When $\ve{\theta}_0\in\Theta_p$, suitable choices of $\mathcal{F}$ and $L$ can lead to an accurate estimate $\hat f_{\mathrm{pre}}(\ve{z}_0)$ of $\ve{\theta}_0$.
However, because $\Theta_p$ is user-specified and may omit plausible regimes, an \emph{out-of-distribution (OOD)} situation may occur if $\ve{\theta}_0\notin\Theta_p$, or if there are not enough samples $\tilde{\ve{\theta}}_i$ in the neighborhood of $\ve{\theta}_0$. Hence, performance can degrade under an OOD, motivating mechanisms that refine $\hat f_{\mathrm{pre}}$ at implementation phase. In this paper, we focus on a special class of simulation-driven estimators $\hat{f}_{\mathrm{pre}}$, known as the TS estimator. 

\section{{The TS Estimator}}\label{sec:ts}
The TS estimator~\cite{garatti2013new} decomposes \eqref{eq:pre_generic} into two-stages:
a fixed \emph{compression} of $\boldsymbol{z}$ into low-dimensional features,
followed by a supervised machine learning model that map these features to the parameters.

\subsection{{Stage 1 (Compression)}}

Choose a deterministic feature map
\begin{equation}\label{eq:compress}
h\colon\mathbb{R}^{2N}\to\mathbb{R}^n,\qquad n\ll 2N,
\end{equation}
e.g., the estimated coefficients of an AR/ARX/ARMAX model fitted to the data.
Let $\ve{x}_i=h(\boldsymbol{z}_i)$ and form the synthetic dataset
\begin{equation}\label{eq:compressed_dataset}
\mathcal{D}_{\mathrm{tr}}
=\{(\ve{x}_i,\tilde{\ve{\theta}}_i)\}_{i=1}^m
=\{(h(\boldsymbol{z}_i),\tilde{\ve{\theta}}_i)\}_{i=1}^m
\subset\mathbb{R}^{n}\times\mathbb{R}^d .
\end{equation}


\subsection{{Stage 2 (Supervised mapping)}}

Let $\ve{x}=h(\mathbf{z})\in\mathbb{R}^n$ be the Stage-1 feature vector.  
The Stage-2 function $g_{\mathrm{pre}}\colon \mathbb{R}^n\to\mathbb{R}^d$ is a fully connected deep neural network with \emph{trunk} and \emph{head} layers.

\subsubsection{{Trunk Layers}}

The trunk of the second stage neural network is mathematically described as follows: Let $\ve{c}^{(0)}=\ve{x}$ be the input layer and for $r=1,\dots,R-1$, the output of the $r^\mathrm{th}$ hidden layer is given by
\(
\ve{c}^{(r)} \;=\; \sigma_r\!\big(\ve{W}_r\, \ve{c}^{(r-1)}\big), 
\,\,
\ve{W}_r\in\mathbb{R}^{p_r\times p_{r-1}},\)
where \(\ve{W}_r\) denotes the weights of the hidden layer $r$, $\sigma_r$ denotes an activation function (e.g. ReLU($\cdot$)), $p_0=n$ and $p_{R-1}=p$. These $R$ layers constitute the \emph{trunk} of $g_{\mathrm{pre}}$. The trunk output is $\ve{t}(\ve{x})\coloneqq \ve{c}^{(R-1)}\in\mathbb{R}^p$ and its parameters are
\(
\ve{\phi} \coloneqq \{\ve{W}_r\}_{r=1}^{R-1}.
\)

\subsubsection{{Head Layers}}


To finally predict the targets $\tilde{\ve{\theta}}_i \in \Theta_p \subseteq \mathbb{R}^d$, which can be multidimensional ($d >1$), we append $V-1$ subsequent hidden layers for each parameter dimension, and these $V-1$ layers corresponding to each individual parameter constitute the \emph{head} of $g_{\mathrm{pre}}$, denoted as $\ve{h}_j^{(v)}$, where \(v =1,\ldots,V-1\). We additionally impose that the weights of the first of these $V-1$ layers are {least correlated}, which is needed to ensure that each parameter is correctly identified. We precisely formalize the structure of the head as the following: 

For each parameter index $j\in\{1,\dots,d\}$, starting from $\ve{h}_j^{(0)}=\ve{t}(\ve{x})$ and for $v=1,\dots,V-1$ define
\(\ve{h}_j^{(v)} \;=\; \sigma_{j,v}\!\big(\ve{U}_{j,v}\, \ve{h}_j^{(v-1)} \big),
\, \ve{U}_{j,v}\in\mathbb{R}^{q_{j,v}\times q_{j,v-1}},
\)
where \(\ve{U}_{j,v}\) corresponds to the weights of hidden layer $v$ of head $j$, and $\sigma_{j,v}$ corresponds to the activation function of $v^{\mathrm{th}}$ hidden layer of head $j$,  with $q_{j,0}=p$.  
The scalar output for head $j$ is linear:
\(
\widehat{{\theta}}_j(\ve{x}) \;=\; \ve{a}_j^\top \ve{h}_j^{(V-1)},
\qquad \ve{a}_j\in\mathbb{R}^{q_{j,V-1}}.
\)
The head parameters are then defined as 
\(
\ve{\psi}_j~\coloneqq\{(\ve{U}_{j,v})_{v=1}^{V-1}, \ve{a}_j\},
\)
and $\ve{\Psi}\coloneqq(\psi_1,\dots,\psi_d)$. Stacking the $d$ heads yields the second stage output as
\begin{equation} \label{eq:g_pre_param}
g_{\mathrm{pre}}(\ve{x};\ve{\phi},\ve{\Psi})\coloneqq
\begin{bmatrix}
\widehat{\theta}_1(\ve{x})\\[-1mm]\vdots\\[-1mm]\widehat{\theta}_d(\ve{x})
\end{bmatrix}
\in\mathbb{R}^d, \, \big[g_{\mathrm{pre}}(\ve{x};\ve{\phi},\ve{\Psi})\big]_j\coloneqq\widehat{\theta}_j(\ve{x}).
\end{equation}

Given the training dataset $\mathcal{D}_{\mathrm{tr}}=\{(\ve{x}_i,\tilde{\ve{\theta}}_i)\}_{i=1}^m$, we
learn $(\ve{\phi},\ve{\Psi})$ by solving the following optimization problem via stochastic gradient descent:
{
\begin{equation} \label{eq:g_param_opt}
\begin{split}
    (\hat{\ve{\phi}},\hat{\ve{\Psi}})\in
\arg\min_{\ve{\phi},\ve{\Psi}}\;
&\frac{1}{m}\sum_{i=1}^m \sum_{j=1}^d
\tilde{w}_j\,l\Big(\widehat{\theta}_j(\ve{x}_i),\, [\tilde{\ve{\theta}}_i]_j\Big)\\
&+\lambda_{\mathrm{orth}}\!\!\sum_{1\le j<k\le d}\!\!\big\|\ve{U}_{j,1}\ve{U}_{k,1}^\top\big\|_F^2,
\end{split}
\end{equation}
where $\tilde{w}_j\ge 0$ are weights chosen as \(
\tilde w_j \;=\; 1 /{\widehat{\mathrm{Var}}([\tilde{\ve{\theta}}]_j)}\), where \(
{\widehat{\mathrm{Var}}([\tilde{\ve{\theta}}]_j)}
\;:=\; m^{-1} \sum_{i=1}^{m}([\tilde{\ve{\theta}}_i]_j-\overline{\theta}_j)^2,
\,
\overline{\theta}_j:= m^{-1} \sum_{i=1}^{m}[\tilde{\ve{\theta}}_i]_j.
\) Here, $l$ is the scalar Huber loss~\cite{ciampiconi2024surveytaxonomylossfunctions}, \([\tilde{{\ve\theta}}_i]_j\) denotes the $j^{\mathrm{th}}$ coordinate of $\tilde{\ve{\theta}}_i$, \(\lambda_{\mathrm{orth}}\ge 0\) is a regularization coefficient, and the last term
encourages diversity among the heads by penalizing the correlations between their first-layer weights. Notice that the overall loss function $L$ in Eq.~\eqref{eq:pre_generic} is related to $l$ as \(
L(\hat{\ve\theta},\ve\theta)
:= \sum_{j=1}^{d} \tilde w_j\,\ell([\hat{\ve\theta}]_j,\,[\ve\theta]_j) + \lambda_{\mathrm{orth}}\!\sum_{1\le j<k\le d}\!\big\|\ve{U}_{j,1}\ve{U}_{k,1}^\top\big\|_F^2.
\)}

The final TS estimator is the composition
\(
\hat f_{\mathrm{pre}}(\boldsymbol{z})
=
g_{\mathrm{pre}}\!\big({h}(\boldsymbol{z});\,\hat{\ve{\phi}},\hat{\ve{\Psi}}\big)
\in\mathbb{R}^d.\)
At implementation phase, given a new trajectory $\boldsymbol{z}_0$ generated by $\mathcal{M}(\ve{\theta}_0)$,
the estimate is
\begin{equation}\label{eq:ts_predict}
\hat{\ve{\theta}}_0 = \hat f_{\mathrm{pre}}(\boldsymbol{z}_0)
=
g_{\mathrm{pre}}\!\big({h}(\boldsymbol{z}_0);\,\hat{\ve{\phi}},\hat{\ve{\Psi}}\big).
\end{equation}

If $\ve{\theta}_0\notin\Theta_p$, then $(\ve{z}_0,\ve{\theta}_0)$ is OOD w.r.t.\ $\mathcal{D}_{\mathrm{tr}}$,
and Eq.~\eqref{eq:ts_predict} may yield a poor estimate of $\ve{\theta}_0$.
We shall now state a fine-tuning approach to overcome this problem.

\section{Fine-Tuning the TS Estimator}
\label{sec: proposed_approach}

We propose a \emph{simulation-driven transfer learning} procedure to improve a pretrained TS estimator under OOD observations. We keep the first stage feature extractor $h$ unchanged and {\emph{fine-tune only the last layers} of the second stage $g_{\mathrm{pre}}(\cdot;\ve{\phi},\ve{\Psi})$: we \emph{freeze} the pretrained $\ve{\phi}$ and \emph{update} only the head layers $\ve{\Psi}$ using a small, targeted synthetic dataset from the DT in a neighborhood of the current TS estimate, following standard transfer learning practice i.e., earlier layers capture generic structure and are preserved, so only the final layers are tuned~\cite{yosinski2014transferable}}. In summary, the procedure consists of the following steps:

\begin{itemize}
  \item \textbf{Step 1 — Feature-space OOD detector.}
  Test whether the observed features $\ve{x}_{\mathrm{obs}}:= {h}(\boldsymbol z_0)$ are statistically consistent with features generated by the simulator at the initial TS prediction
  $\hat{\ve{\theta}}_{\mathrm{init}} = g_{\mathrm{pre}}\!\big({h}(\boldsymbol z_0);\hat{\ve{\phi}},\hat{\ve{\Psi}}\big)$.
  If consistent, skip fine-tuning and deploy $\hat{{f}}_\text{pre}$.

  \item \textbf{Step 2 — Synthetic data generation.}
  If an OOD is detected, perform a few Gauss-Newton (GN) iterations in feature space, regularized toward $\hat{\ve{\theta}}_{\mathrm{init}}$, to obtain an improved estimate $\hat{\ve{\theta}}_{\mathrm{GN}}$.
  Then, derive a confidence region around $\hat{\ve{\theta}}_{\mathrm{GN}}$; sample an informative cloud of parameters from this region; and simulate the DT at these samples to form a small synthetic dataset of feature–parameter pairs.

  \item \textbf{Step 3 — Final-layer transfer.}
  With $\ve{\phi}$ frozen, update only $\ve{\Psi}$ on the small synthetic dataset using a weighted loss. The result is an improved estimator
  $g_{\mathrm{pre}}(\cdot;\hat{\ve{\phi}},\hat{\ve{\Psi}})$ under the OOD.
\end{itemize}


\subsection{Step 1: Feature–Space OOD Detector}
\label{subsec:gate}

{

We now provide a procedure to detect an OOD scenario. To this end, we appeal to a bootstrapped version of Wald's test~\cite{wald1943tests,wasserman2013all,greene2003econometric}. The null hypothesis, \(H_0\), is that the observed features, \(\ve{x}_{\mathrm{obs}}\), are statistically consistent with the features generated by the simulator when its parameter is set to the initial estimate, \(\hat{\ve{\theta}}_{\mathrm{init}}\coloneqq g_{\mathrm{pre}}\!\big(\ve{x}_{\mathrm{obs}};\hat{\ve{\phi}},\hat{\ve{\Psi}}\big)\).

The procedure begins by computing the initial parameter estimate from the new observations, \(\boldsymbol z_0\), generated by the system $\mathcal{M}(\ve{\theta}_0)$, where $\ve{\theta}_0$ is unknown. This is done by first extracting its features, \(\ve{x}_{\mathrm{obs}}:=h(\boldsymbol z_0)\), and then applying the pre-trained second-stage estimator, outputting initial estimate of $\ve{\theta}_0$ as
\(
\hat{\ve{\theta}}_{\mathrm{init}} := g_{\mathrm{pre}}\!\big(\ve{x}_{\mathrm{obs}};\hat{\ve{\phi}},\hat{\ve{\Psi}}\big).
\)
Next, we fix the simulator's parameter at \(\hat{\ve{\theta}}_{\mathrm{init}}\) and run \(K\) simulations, each with a different random seed \(\omega_k\). Applying the feature extractor \({h}\) to each of the \(K\) resulting trajectories yields a \(K \times n\) matrix of simulated features:
\(
\ve{H} \coloneqq
(
(\ve{x}^{(1)})^\top, \ldots,(\ve{x}^{(K)})^\top)^\top
,
\)
{where $\ve{x}^{(k)} = h(\ve{z}_k)$, and $\ve{z}_k$ is the input–output trajectory generated by the simulator $\mathcal{M}(\hat{\ve{\theta}}_{\mathrm{init}})$ at noise seed $\omega_k$, for $k=1,\ldots,K$.} 
This matrix, \(\ve{H}\), characterizes the expected variability of the features if the true parameter were indeed \(\hat{\ve{\theta}}_{\mathrm{init}}\). To construct a test statistic, we compute a whitened discrepancy measure. From the simulated features \(\ve{H}\) (at parameter \(\hat{\ve{\theta}}_{\mathrm{init}}\)), estimate the feature mean and regularized covariance as functions of \(\ve{\theta}\):
\begin{equation}\label{eq:simulated_features}
\begin{split}
    \ve{\mu}&:=\frac{1}{K}\sum_{k=1}^K \ve{H}_{k:}^\top, \\
    \ve{S} &:=\frac{1}{K-1}\sum_{k=1}^K \big[\ve{x}^{(k)}-\ve{\mu}\big] \big[\ve{x}^{(k)}-\ve{\mu}\big]^\top \\
    &\qquad\qquad\qquad\qquad +\varepsilon \ve{I}_n,\;\varepsilon>0,
\end{split}
\end{equation}
where \(\ve{H}_{k:}=(\ve{x}^{(k)})^\top\), and $\ve{I}_n$ is an $n \times n$ identity matrix. The covariance matrix is then used to define a whitening transform, \(\ve{S}^{-1/2}\), which normalizes the feature space by accounting for correlations and different scales among the features. The test statistic for our observed data is the whitened squared residual norm:
\begin{equation}
s_{\mathrm{obs}}=\big\|\ve{S}^{-1/2}\big(\ve{x}_{\mathrm{obs}}-\ve{\mu}\big)\big\|_2^2,
\end{equation}

Finally, we establish a critical value for this statistic using the bootstrap principle. We compute the same whitened statistic for each of the \(K\) simulated feature vectors:
\begin{equation}
s_k=\big\|\ve{S}^{-1/2}\big(\ve{H}_{k:}^\top-\ve{\mu}\big)\big\|_2^2.
\end{equation}
The set \(\{s_k\}_{k=1}^K\) provides an empirical distribution of the test statistic under \(H_0\). We define our decision threshold, \(q_{1-\alpha}\), as the empirical \((1-\alpha)\)-quantile of this distribution.

\textbf{Decision rule:} If \(s_{\mathrm{obs}} > q_{1-\alpha}\), we reject the null hypothesis $H_0$, conclude that an OOD scenario is likely, and proceed to Step~2. Otherwise, the observation is deemed statistically consistent with the variability at \(\hat{\ve{\theta}}_{\mathrm{init}}\), and we retain the pretrained estimator without fine-tuning.

}
\subsection{Step 2: Synthetic data generation}
\label{subsec:align_fisher}
{
If \(s_{\mathrm{obs}} > q_{1-\alpha}\), then the {compressed features from the first-stage of TS} evaluated at \(\hat{\ve{\theta}}_{\mathrm{init}}\) fail to explain \(\ve{x}_{\mathrm{obs}}\) in the whitened space. In that case, we refine \(\hat{\ve{\theta}}_{\mathrm{init}}\) by running GN iterations that minimize the whitened feature residual between the observed feature and the first-stage generated feature. Because this residual is random, the GN estimate, denoted \(\hat{\ve{\theta}}_{\mathrm{GN}}\), still carries parameter uncertainty. To effectively fine-tune the second stage of the TS estimator, we need to generate a cloud of synthetic data that covers the region of plausible parameter values around this estimate. We shall now describe precisely the synthetic dataset generation for fine-tuning the second stage of TS as follows.

\paragraph{GN update}
We compute the GN update starting from the TS initial estimate \(\hat{\ve{\theta}}_{\mathrm{init}}\). We define a seed-averaged simulator feature map \(\bar{\ve{f}}(\ve{\theta})\) and its corresponding whitened residual \(\ve{r}(\ve{\theta})\):
\begin{equation}\label{eq:bar_f}
\begin{aligned}
\bar{\ve{f}}(\ve{\theta})
&:= \frac{1}{K_{\mathrm{avg}}}\sum_{\omega\in\mathcal{S}_{\mathrm{avg}}}
{h}\!\big(\boldsymbol z(\ve{\theta},\omega;\ve{u})\big),\\
\ve{r}(\ve{\theta})
&:= \ve{S}^{-1/2}\big(\bar{\ve{f}}(\ve{\theta})-\ve{x}_{\mathrm{obs}}\big)\in\mathbb{R}^{n},
\end{aligned}
\end{equation}
where \(\ve{z}(\ve{\theta},\omega;\ve{u})\) is the simulator trajectory for parameter \(\ve{\theta}\) and seed \(\omega\). We then minimize the following regularized objective function using iterative GN steps (with a backtracking line search):
\begin{equation}
v(\ve{\theta})=\tfrac{1}{2}\|\ve{r}(\ve{\theta})\|_2^2+\tfrac{\gamma}{2}\|\ve{\theta}-\hat{\ve{\theta}}_{\mathrm{init}}\|_2^2,\qquad \gamma>0,
\label{eq:gn-obj}
\end{equation}
A few iterations of minimizing this objective yield the refined estimate \(\hat{\ve{\theta}}_{\mathrm{GN}}\). Let \(\partial \bar{\ve{f}}(\ve{\theta})/\partial \ve{\theta} \in \mathbb{R}^{n\times d}\) denote the Jacobian of the seed-averaged feature map, and let \(\ve{S}(\ve{\theta})\) be the feature covariance estimated from simulator draws {at parameter \(\ve{\theta}\)}, analogous to Eq.~\eqref{eq:simulated_features}. Define the {whitened Jacobian~
\(
\ve{J} \;\coloneqq\; \ve{S}^{-1/2}\, \partial \bar{\ve{f}}(\ve{\theta})/\partial \ve{\theta}.
\)} We re-compute \(\ve{\mu}, \ve{S}\) and \(\ve{J}\) at $\hat{\ve{\theta}}_{\mathrm{GN}}$, and set
\(
\ve{G} \coloneqq \ve{J}^\top \ve{J} + \lambda \ve{I}_d.
\)

\paragraph{Sampling procedure}
\(\hat{\ve{\theta}}_{\mathrm{GN}}\) is a point estimate derived from a single, noisy observation and is therefore uncertain. To effectively fine-tune the TS estimator, we must generate a cloud of synthetic data that explores the region of plausible parameter values around this estimate. We employ a hybrid sampling strategy that combines a \emph{confidence-ellipsoid} based method with an alternative sampling scheme.

Our primary approach is to sample from the approximate \(1-\beta\) {confidence ellipsoid} for the true parameter \(\ve{\theta}_0\). This region is defined by the set of all \(\ve{\theta}\) satisfying:
\begin{equation}
    (\ve{\theta}-\hat{\ve{\theta}}_{\mathrm{GN}})^\top \ve{G} (\ve{\theta}-\hat{\ve{\theta}}_{\mathrm{GN}}) \leq \chi^2_d(\beta),
\end{equation}
where \(\chi^2_d(\beta)\) is the chi-squared quantile and \(\ve{G}\) is the approximate Fisher Information Matrix (FIM). This ellipsoid represents a high-confidence region, and sampling from it is statistically efficient for exploring well-identified parameters. However, this primary method can fail if the model is insensitive to one or more parameters. In such cases, the FIM, \(\ve{G}\), becomes ill-conditioned, causing the confidence ellipsoid to be stretched along the insensitive directions, and thereby making the sampling procedure numerically unstable. To overcome this, we alternate between this sampling approach and \emph{sensitivity} sampling. The latter sampling procedure handles the ill-conditioned directions. It works by generating samples along individual parameter axis with a step size that is inversely proportional to the simulator's sensitivity to that parameter. The sensitivity of coordinate \(j\) is defined as \(s'_j := \|\ve{J} \ve{e}_j\|_2\). To set the sampling scale, we define two quantities. First, \(\Delta_{\mathrm{misfit}}:=\|\ve{r}(\hat{\ve{\theta}}_{\mathrm{GN}})\|_2\), measures the discrepancy between the observed summary statistic and the one generated by the simulator at the point estimate \(\hat{\ve{\theta}}_{\mathrm{GN}}\). Second, we compute \(d_{\mathrm{med}}:=\mathrm{median}_k\, d_k\), based on the spread of pre-computed summary features, where each \(d_k:=\|\ve{S}^{-1/2}(\ve{H}_{k:}-\ve{\mu})\|_2\) represents the whitened residual. We then set coordinate scales, \(r_j\), that increase exploration for less sensitive parameters (small \(s^{\prime}_j\)) while also adapting to the simulator's noise level and the current feature mismatch as
\(r_j \;=\; \frac{d_{\mathrm{med}}}{s^{\prime}_j}\,\Delta_{\mathrm{misfit}}.
\label{eq:rj-floor}
\)

Our sampling procedure combines confidence ellipsoid and sensitivity sampling, yielding a hybrid sampling scheme, given by
\(\bar{\ve{\theta}}_k=\hat{\ve{\theta}}_{\mathrm{GN}}+\ve{\Delta\theta}_k\), where $\ve{\Delta \theta}_k$ is drawn as
\begin{align}\label{eq:proposal_mixer}
&\text{(confidence set sampling)}\quad
\ve{\Delta\theta}_k = \ve{B}\,\ve{z},\;\; \ve{z}\sim\mathcal{N}(\ve{0},\ve{I}_d),\;\; \nonumber \\
&\quad \quad \qquad \qquad \qquad \qquad \qquad \qquad \quad \ve{B}\ve{B}^\top \approx \ve{G}^{-1}, \nonumber\\
&\text{(sensitivity sampling)}\,
\text{pick } i\sim\mathrm{Unif}\{1,\ldots,d\},\; \zeta\sim\mathcal{N}(0,r_i^2), \nonumber\\
&\quad \qquad \qquad \qquad \qquad \, \, \quad \quad \ve{\Delta\theta}_k=\zeta\,\ve{e}_i,
\end{align}
where \(\ve{e}_i \in \mathbb{R}^d \) is the $i$-th standard basis vector ($1$ at position $i$, zeros elsewhere), and $\ve{B}$ is the Cholesky factor of $\ve{G}^{-1}$.
To create the {fine-tuning dataset}, \(\mathcal{D}_{\mathrm{fine}}\), we generate a synthetic observation \(\bar{\ve{z}}_k\) from the simulator for each parameter sample \(\bar{\ve{\theta}}_k\). Each observation is then compressed using \(h\), forming a dataset of pairs: \(\mathcal{D}_{\mathrm{fine}} = \{(h(\ve{\bar{z}}_k),\bar{\ve{\theta}}_k)\}_{k=1}^{M_{\mathrm{FT}}}\), {where \(M_{\mathrm{FT}}\) is the number of new synthetic samples.}

\begin{remark}
The quantity~\(\Delta_{\mathrm{misfit}}\) can be artificially small due to noise induced by the randomness of the simulator $\mathcal{M}$, so the naive radius \(\Delta_{\mathrm{misfit}}/s^{\prime}_j\) would under–explore.
Multiplying by
\(d_{\mathrm{med}}\)
rescales the step to the \emph{typical} feature fluctuation at \(\hat{\ve{\theta}}_{\mathrm{init}}\), giving
\(r_j=(d_{\mathrm{med}}/s^{\prime}_j)\,\Delta_{\mathrm{misfit}}\).
This preserves the desired inverse–sensitivity (\(1/s^{\prime}_j\)) while making the samples noise–aware.
\end{remark}
}

{
\begin{remark}[Theoretical justification of Steps 1 and 2]
{For Step 1,} under the null hypothesis (in-distribution), the observed features \(\ve{x}_{\mathrm{obs}}\) and the simulated features \(\ve{x}^{(k)}\) are asymptotically normal with the same mean and covariance. This implies that their whitened residuals (\(s_{\mathrm{obs}}\) and \(s_k\)) converge to the same \(\chi^2\) distribution, validating the bootstrap test.
{For Step 2,} the GN objective \(v(\ve{\theta})\) is an M-estimator whose expectation is minimized at the true \(\ve{\theta}_0\). Therefore, the GN estimate \(\hat{\ve{\theta}}_{\mathrm{GN}}\) is consistent~\cite[Ch. 5]{van2000asymptotic}, and the FIM \(\ve{G}\), as a consistent estimator of the asymptotic covariance, defines an asymptotically valid confidence ellipsoid for sampling.
\end{remark}}


\subsection{Step 3: Final–Layer Transfer on a Targeted Set}
\label{subsec:head}
Finally, we freeze the early layers of $g_{\mathrm{pre}}(\cdot;\ve{\phi},\ve{\Psi})$ (i.e., keep $\ve{\phi}$ fixed) and modify
only the last layers, yielding $g_{\mathrm{ft}}(\cdot;\ve{\phi},\ve{\Psi}_{\mathrm{ft}})$. We fit $\ve{\Psi}$ on $\mathcal D_{\mathrm{fine}}$ with a small number of gradient steps
using a weighted objective (cf.~\eqref{eq:g_param_opt}):
{
\begin{equation} \label{eq:ft-obj}
\begin{aligned}
    \hat{\ve{\Psi}} \in \arg \min_{\ve{\Psi}}\;
\frac{1}{M_{\mathrm{FT}}}\sum_{i=1}^{M_{\mathrm{FT}}}&\sum_{j=1}^d
\tilde{w}_j
\ell([g_{\mathrm{pre}}(h(\bar{\ve{z}}_i);\ve{\phi},\ve{\Psi})]_j,[\bar{\ve{\theta}}_{i}]_j)
\\
&+\lambda_{\mathrm{orth}}\!\!\sum_{1\le j<k\le d}\!\!\big\|\ve{U}_{j,1}\ve{U}_{k,1}^\top\big\|_F^2,
\end{aligned}
\end{equation}
where \(
\tilde w_j
\;=\;
d\, [\ve{G}]_{jj} / \sum_{k=1}^{d} [\ve{G}]_{kk} \), with $[\ve{G}]_{jj}$ denoting the $j$-th diagonal entry of $\ve{G}$, and $\ell$ is the Huber loss function.}

\section{Simulation Study}
\label{sec:simulation_study}

We evaluate the proposed fine-tuning scheme on two nonlinear systems:
(i) the Van der Pol (VDP) oscillator and (ii) cascaded water tanks. In both cases,
the model structure is known but the true parameters may lie outside the sampling
set used to pretrain the TS estimator. We compare the fine-tuned TS (\textbf{TS-fine}) against
pretrained TS (\textbf{TS-pre}), the PEM,
and a dual-EKF. For PEM and dual-EKF we report
results under three initializations: \emph{Good Init} (a Gaussian perturbation of the true parameter value), \emph{TS-Init} (initialized at
\(\hat f_{\mathrm{pre}}(\boldsymbol z_0)\)), and \emph{Bad/Worse-Case Init } (point far from
\(\ve{\theta}_0\)).

\paragraph{Common settings}
For all simulations we set \(K=150\) in~\eqref{eq:simulated_features}.
The whitened Jacobian \(\ve{J}\) is obtained via finite differences.
To compute \(\bar {\ve{f}}(\ve{\theta})\) in~\eqref{eq:bar_f} and the Jacobians,
we use common random numbers with the seed set
\(S_{\mathrm{avg}}=\{8000,8001,8002\}\).
Both examples are simulated in discrete time using forward Euler with a fixed step size
\(\Delta t\) (VDP: \(0.05\) s; tanks: \(0.5\) s).
The Gauss–Newton regularization parameter in~\eqref{eq:gn-obj} is
\(\gamma=10^{-3}\) (VDP) and \(\gamma=10^{-4}\) (tanks).
The orthogonality regularizer $\lambda_{\mathrm{orth}}$ on the first head layer is
\(0\) for VDP (single output) and \(5\times10^{-4}\) for the tanks.

\paragraph{Feature compression and stage-2 network.}
For VDP we use \(\mathrm{AR}(5)\) coefficients fitted to \(y\); for the tanks we use
\(\mathrm{ARX}(64,64)\) coefficients fitted to \((y,u)\).
The stage-2 regressor \(g_{\mathrm{pre}}\) is a fully connected network with a shared trunk
and one head per parameter, and the following settings:
\begin{itemize}
  \item \textbf{VDP:} trunk \((256,128)\) with ReLU+Dropout(0.1);
        one head (depth 2, hidden 128, Tanh+Dropout(0.1)).
        Adam with lr \(10^{-3}\), batch size \(64\), \(200\) epochs.
        Offline bank size \(m=10000\); fine-tune cloud \(M_{\mathrm{FT}}=400\).
  \item \textbf{Tanks:} trunk \((256,256)\) with LayerNorm+LeakyReLU+Dropout(0.15);
        four heads (depth 3, hidden 64, LayerNorm+Tanh+Dropout(0.15)).
        Adam with lr \(10^{-3}\), batch size \(64\), \(250\) epochs.
        Offline bank size \(m=10000\); fine-tune cloud \(M_{\mathrm{FT}}=1000\).
\end{itemize}
In the fine-tuned TS estimator (TS-fine), we {freeze the trunk} and {retrain only the head layers} on the synthetic cloud around the GN update; for the tanks we also use a smaller head
(hidden 64, depth 1, dropout 0.20) with early stopping.

{\begin{remark}
We fix all hyperparameters ($K$, $M_{\mathrm{FT}}$, $\gamma$, $\lambda_{\mathrm{orth}}$, architectures, and training choices) \emph{a priori} across all experiments.
\end{remark}}

\subsection{Van der Pol Oscillator}
\label{subsec:vdp}
The VDP dynamics are
\begin{equation}\label{eq:vdp}
  \dot x_1 = x_2,\qquad
  \dot x_2 = \nu(1-x_1^2)x_2 - x_1 + w_k,\qquad
  y=x_1,
\end{equation}
with unknown \(\nu>0\) and small process noise \(w_k\sim\mathcal N(0,0.02)\).
We collect \(N=300\) length observations for each \(\nu_i\sim\mathrm{Unif}[0,2.5]\) to form the
offline training set \(\mathcal D_{\mathrm{tr}}\). 

\begin{figure}[h!]
  \centering
    \includegraphics[width=1.0\linewidth]{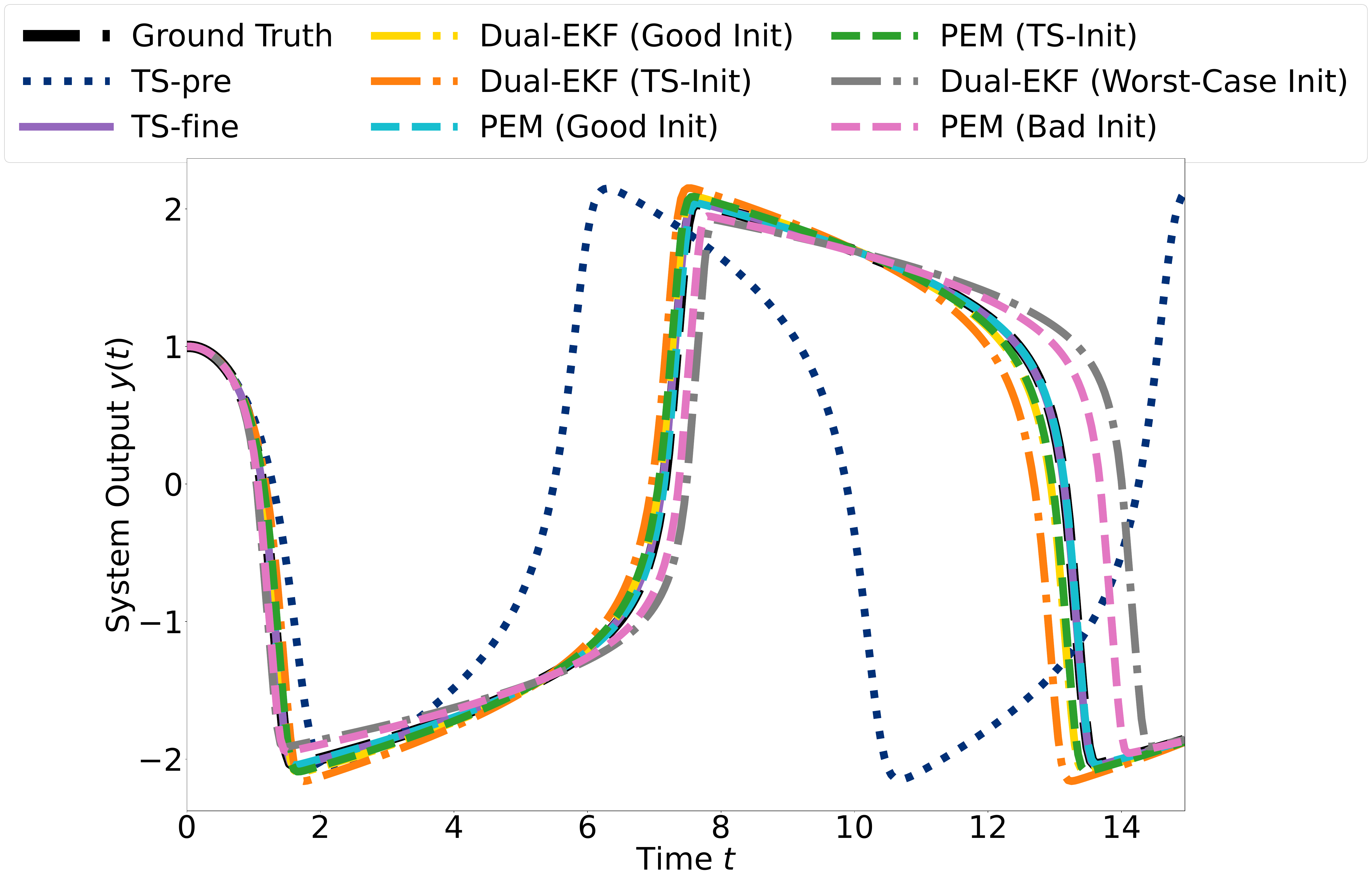}
    \caption{VDP (OOD 2): ground-truth vs.\ mean predictions over $100$ MC runs. The trajectory plots are measurements without noise.}
    \label{fig:vdp_traj_o3}
\end{figure}

\textbf{OOD tests.}
We evaluate at \(\nu=3.0\) (OOD $1$), {a mild OOD} and \(\nu=5.0\) (OOD $2$), which are outside the pretraining range.
Over \(n_{\mathrm{mc}}=100\) Monte-Carlo (MC) runs, we report mean squared errors (MSEs) for OOD $1$ and average trajectories of the
true system~\eqref{eq:vdp} and of each method for OOD $2$ (Figs.~\ref{fig:vdp_traj_o3} and Table~\ref{tab:ood})\footnote{{We report OOD\,1 quantitatively (Table~\ref{tab:ood}) and OOD\,2 qualitatively (Figs.~\ref{fig:vdp_traj_o3}, \ref{fig:tanks_traj_o2}), since both OOD settings show the same relative gaps, so we present complementary views.}}.
TS-fine has a substantially smaller MSE than TS-pre and outperforms dual-EKF/PEM under poor initializations; PEM with a good initialization can perform well, but is initialization-sensitive, unlike TS-fine.

\begin{remark}
     We set \(\alpha \in [0.05,\,0.10]\).This range was selected from a preliminary sensitivity analysis to balance false positives on in-distribution data with reliable OOD detection. 
Unless otherwise noted, we use \(\alpha=0.10\) in all experiments.
\end{remark}

\subsection{Cascaded Water Tanks}
\label{subsec:tanks}
The cascaded water tanks dynamics are given by
\begin{equation}
\begin{aligned}
  \dot h_1 &= -k_1\sqrt{|h_1|} + k_4\,u_k + w^{(1)}_k,\\
  \dot h_2 &= \phantom{-}k_2\sqrt{|h_1|} - k_3\sqrt{|h_2|} + w^{(2)}_k,\\
  y_k &= h_{2,k} + v_k,
\end{aligned}
\end{equation}
with parameter vector \(\ve{\theta}=(k_1,k_2,k_3,k_4)^\top\),
process noises \(w^{(1)}_k,w^{(2)}_k\stackrel{\text{i.i.d.}}{\sim}\mathcal N(0,\,0.01)\),
and measurement noise \(v_k\sim\mathcal N(0,\,0.02)\).
We use \(\Delta t=0.5\) s, and measure \(y_k\) as output.

\textbf{Offline pretraining set.}
We take samples \(\tilde{\ve{\theta}}_i\sim [0.4,0.8] \times [0.4,0.8] \times [0.4,0.8]\times[0.8,1.2]\) uniformly (\(m=10000\)) and for each $\tilde{\ve{\theta}}_i$, we collect $N=400$ length observations.
The inputs are PRBS signals~\cite{wigren2013three}: the level of binary signal is drawn i.i.d. from \([0,3]\)
and held for \(30\) samples.

\textbf{OOD tests.}
We consider \(\ve{\theta}=(1.2,1.2,0.9,1.0)^\top\)
 (OOD $1$) and \(\ve{\theta}=(1.3,1.3,0.6,0.7)^\top\)~(OOD $2$), both outside the pretraining set. For evaluation we fix the same block-uniform input across runs, simulate with the noise model above, and average over \(n_{\mathrm{mc}}=100\) MC runs. A comparison of the results in terms of MSE and trajectory tracking is presented in Table~\ref{tab:ood} for OOD $1$ and Fig.~\ref{fig:tanks_traj_o2}. From Fig.~\ref{fig:tanks_traj_o2} and Table~\ref{tab:ood}, we observe that TS-fine improves upon TS-pre and is competitive with
(or better than) dual-EKF/PEM without relying on any initialization.
PEM can excel with a good initialization; TS-fine is initialization-free.

\section{{Future work}}\label{sec:future}
{The following are possible directions for future work:
\begin{enumerate}
    \item The accuracy of TS-pre and TS-fine depends on both stages. Under high process/measurement noise (low SNR), the Stage~1 compressor should yield statistics from which the parameters can be inferred {irrespective of the noise level}, while Stage~2 (MLP) can become unreliable during training. Our aim here is to fine-tune TS, not to optimize the underlying first/second-stage architectures for noise robustness. Consequently, in low SNR regimes, the standard baselines may outperform our current setup. This can be addressed by (a) designing a noise-robust Stage~1 compressor and (b) strengthening Stage~2 via adversarial training.
    \item Fine-tuning adds a computation overhead from three sources: (i) computing $\ve{J}$ and solving the $d\times d$ system $\ve{G}=\ve{J}^\top \ve{J}+\lambda \ve{I}_d$, which has $\mathcal{O}(d^3)$ time complexity, (ii) generating a synthetic cloud of $M_{\mathrm{FT}}$ parameter samples and their features, and (iii) short, head-only retraining over this cloud. This computation overhead can be overcome via \emph{unsupervised} fine-tuning that updates only the final-layer weights by directly shrinking the discrepancy between simulated and observed trajectories/features, thereby avoiding large synthetic clouds.
\end{enumerate}
}

{
\begin{table}[h]
\centering
\caption{OOD 1 results: MSE and avg. computation time (ms).}
\label{tab:ood}
\begingroup
\scriptsize
\setlength{\tabcolsep}{3pt}
\resizebox{\linewidth}{!}{
\begin{tabular}{
l
S[round-mode=places,round-precision=6] S[round-mode=places,round-precision=3]
S[round-mode=places,round-precision=6] S[round-mode=places,round-precision=3]
}
\toprule
& \multicolumn{2}{c}{\textbf{VDP OOD 1}} & \multicolumn{2}{c}{\textbf{Tanks OOD 1}} \\
\cmidrule(lr){2-3}\cmidrule(lr){4-5}
\textbf{Method} & {\textbf{MSE}} & {\textbf{ms}} & {\textbf{MSE}} & {\textbf{ms}} \\
\midrule
TS-pre                  & 0.3204    & 1.885488 & 0.1189   & 3.994949 \\
TS-fine                 & 0.0005501 & 13087.631394 & 0.07051  & 50854.723691 \\
Dual-EKF (GI)           & 0.003269  & 8.289987 & 6.112    & 18.850671 \\
Dual-EKF (TI)           & 0.00673   & 8.401977 & 8.345    & 19.106957 \\
PEM (GI)                & 0.001154  & 138.949625 & 0.07367  & 8564.443639 \\
PEM (TI)                & 0.001154  & 115.814819 & 0.186    & 8783.367652 \\
Dual-EKF (WI)           & 15.78     & 8.368522 & 8.978    & 19.033702 \\
PEM (BI)                & 14.67     & 83.472527 & 0.3924   & 7086.184831 \\
\bottomrule
\end{tabular}
}
\par\medskip
\raggedright\emph{Init:} GI = Good, TI = TS-init, WI = Worst-case Init, BI = Bad Init.\par
\endgroup
\end{table}


\begin{figure}[h!]
\centering
\includegraphics[width=1\linewidth]{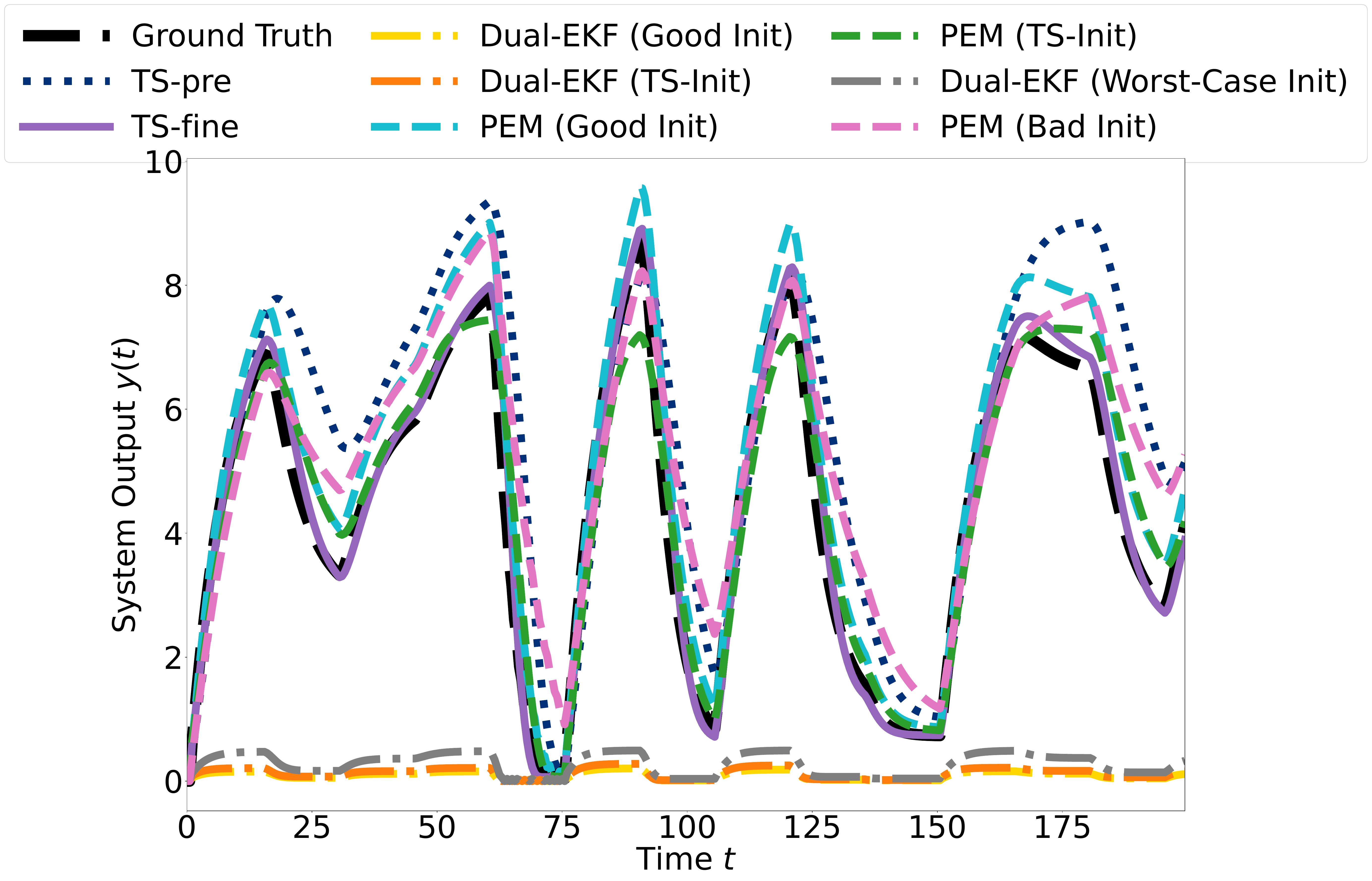}
\caption{Tanks (OOD 2): average trajectory (ground truth vs.\ mean predictions over $100$ MC runs. The trajectory plots are measurements without noise.}
\label{fig:tanks_traj_o2}
\end{figure}
}


\section{Conclusion} \label{sec: conclusion}

In this paper, we have proposed a transfer-learning approach to fine-tune a pretrained, simulation-driven (TS) estimator for OOD scenarios. We first apply a statistical detection test on the compressed features to flag OOD; if the test detects an OOD, we modify only the second stage, a neural network, by freezing the shared trunk and retraining the last, per-parameter layers using a newly synthesized local dataset. Numerical examples demonstrate that the fine-tuned TS delivers improved performance over the pretrained TS and fares well against initialization-dependent baselines such as dual-EKF and PEM.

\bibliographystyle{IEEEtran}
\bibliography{ref}

@preamble{ " \newcommand{\noop}[1]{} " }

@article{garatti2013new,
  title={A new paradigm for parameter estimation in system modeling},
  author={S. Garatti and S. Bittanti},
  journal={International Journal of Adaptive Control and Signal Processing},
  volume={27},
  number={8},
  pages={667--687},
  year={2013},
}

@book{van2000asymptotic,
  title={Asymptotic Statistics},
  author={Van der Vaart, A. W.},
  year={2000},
  publisher={Cambridge University Press}
}

@article{GHOSH2024111327,
title = {{DeepBayes}--{An} estimator for parameter estimation in stochastic nonlinear dynamical models},
author = {A. Ghosh and M. Abdalmoaty and S. Chatterjee and H. Hjalmarsson},
journal = {Automatica},
volume = {159},
pages = {111327},
year = {2024}
}

@article{vanderhorn2021digital,
  title={Digital Twin: Generalization, characterization and implementation},
  author={VanDerHorn, E. and Mahadevan, S.},
  journal={Decision support systems},
  volume={145},
  pages={113524},
  year={2021},
  publisher={Elsevier}
}

@ARTICLE{Diskin,
  author={Diskin, T. and Eldar, Y. C. and Wiesel, A.},
  journal={IEEE Transactions on Signal Processing}, 
  title={Learning to Estimate Without Bias}, 
  year={2023},
  volume={71},
  number={},
  pages={2162-2171},
  doi={10.1109/TSP.2023.3284372}}

@inproceedings{dettu2024data,
  title={From data to control: a two-stage simulation-based approach},
  author={Dett{\`u}, F. and Lakshminarayanan, B. and Formentin, S. and Rojas, C. R.},
  booktitle={Proceedings of the 2024 European Control Conference (ECC)},
  pages={3428--3433},
  year={2024}
}

@article{kritzinger2018digital,
  title={Digital Twin in manufacturing: A categorical literature review and classification},
  author={Kritzinger, W. and Karner, M. and Traar, G. and Henjes, J. and Sihn, W.},
  journal={IFAC-PapersOnline},
  volume={51},
  number={11},
  pages={1016--1022},
  year={2018},
}

@article{Grieves2017,
  title={Digital twin: Mitigating unpredictable, undesirable emergent behavior in complex systems},
  author={M. Grieves and J. Vickers},
  journal={Transdisciplinary perspectives on complex systems: New findings and approaches},
  pages={85--113},
  year={2017}
}

@article{forgione2023adaptation,
  title={On the adaptation of recurrent neural networks for system identification},
  author={Forgione, M. and Muni, A. and Piga, D. and Gallieri, M.},
  journal={Automatica},
  volume={155},
  pages={111092},
  year={2023},
  publisher={Elsevier}
}

@article{niu2022deep,
  title={Deep transfer learning for system identification using long short-term memory neural networks},
  author={Niu, K. and Zhou, M. and Abdallah, C. T. and Hayajneh, M.},
  journal={arXiv preprint arXiv:2204.03125},
  year={2022}
}

@article{pan2009survey,
  title={A survey on transfer learning},
  author={Pan, S. J. and Yang, Q.},
  journal={IEEE Transactions on knowledge and data engineering},
  volume={22},
  number={10},
  pages={1345--1359},
  year={2009},
  publisher={IEEE}
}

@inproceedings{wigren2013three,
  title={Three free data sets for development and benchmarking in nonlinear system identification},
  author={Wigren, T. and Schoukens, J.},
  booktitle={2013 European control conference (ECC)},
  pages={2933--2938},
  year={2013},
  organization={IEEE}
}

@article{lakshminarayanan2025asymptotic,
  title={On Asymptotic Analysis of the Two-Stage Approach: Towards Data-Driven Parameter Estimation},
  author={Lakshminarayanan, B. and Rojas, C. R.},
  journal={arXiv preprint arXiv:2508.18201},
  year={2025}
}

@article{wald1943tests,
  title={Tests of statistical hypotheses concerning several parameters when the number of observations is large},
  author={Wald, A.},
  journal={Transactions of the American Mathematical society},
  volume={54},
  number={3},
  pages={426--482},
  year={1943},
  publisher={JSTOR}
}

@book{wasserman2013all,
  title={All of statistics: a concise course in statistical inference},
  author={Wasserman, L.},
  year={2013},
  publisher={Springer Science \& Business Media}
}

@article{wan2001dual,
  title={Dual extended {Kalman} filter methods},
  author={Wan, E. and Nelson, A.},
  journal={Kalman filtering and neural networks},
  pages={123--173},
  year={2001},
  publisher={Wiley Online Library}
}

@article{yosinski2014transferable,
  title={How transferable are features in deep neural networks?},
  author={Yosinski, J. and Clune, J. and Bengio, Y. and Lipson, H.},
  journal={Advances in neural information processing systems},
  volume={27},
  year={2014}
}

@article{greene2003econometric,
  title={Econometric analysis},
  author={Greene, W.},
  journal={Pretence Hall},
  year={2003}
}

@inproceedings{piga2024synthetic,
  title={Synthetic data generation for system identification: leveraging knowledge transfer from similar systems},
  author={Piga, D. and Rufolo, M. and Maroni, G. and Mejari, M. and Forgione, M.},
  booktitle={2024 IEEE 63rd Conference on Decision and Control (CDC)},
  pages={6383--6388},
  year={2024},
  organization={IEEE}
}

@book{pml1Book,
 author = "K. P. Murphy",
 title = "Probabilistic Machine Learning: An introduction",
 publisher = "MIT Press",
 year = 2022
}

@article{zeghal2022neural,
  title={Neural posterior estimation with differentiable simulators},
  author={Zeghal, J. and Lanusse, F. and Boucaud, A. and Remy, B. and Aubourg, E.},
  journal={arXiv preprint arXiv:2207.05636},
  year={2022}
}

@article{deistler2022truncated,
  title={Truncated proposals for scalable and hassle-free simulation-based inference},
  author={Deistler, M. and Goncalves, P. J. and Macke, J. H. },
  journal={Advances in neural information processing systems},
  volume={35},
  pages={23135--23149},
  year={2022}
}

@misc{ciampiconi2024surveytaxonomylossfunctions,
      title={A survey and taxonomy of loss functions in machine learning}, 
      author={L. Ciampiconi and A. Elwood and M. Leonardi and A. Mohamed and A. Rozza},
      year={2024},
      eprint={2301.05579},
      archivePrefix={arXiv},
      primaryClass={cs.LG}
}

@article{ward2022robust,
  title={Robust neural posterior estimation and statistical model criticism},
  author={Ward, D. and Cannon, P. and Beaumont, M. and Fasiolo, M. and Schmon, S.},
  journal={Advances in Neural Information Processing Systems},
  volume={35},
  pages={33845--33859},
  year={2022}
}

\end{document}